# Modeling resonance characteristics of the Chang'e-7 lander modulated by solar panel rotation under lunar south-pole thermal environment

**Lei Zhang[1], Jinhai Zhang[1]***

[1] Center for Deep Earth Technology and Equipment, Key Laboratory of Deep Petroleum Intelligent Exploration and Development, Institute of Geology and Geophysics, Chinese Academy of Sciences, Beijing 100029, China.

Corresponding author: Jinhai Zhang (zjh@mail.iggcas.ac.cn)

**Key Points:**

- FEM simulations characterize CE-7 lander resonance under coupled solar panel rotation and extreme polar thermal cycles.
- Overlapping the seismic band, the 0.76 Hz mode drifts between 0.64–0.87 Hz due to thermal stiffness variations.
- Sensitivity analysis identifies the supporting bracket, rather than the solar panel, as the primary stiffness bottleneck driving drift.

**Plain Language Summary**

The Chang'e-7 mission will land in the Moon's south pole and deploy a seismometer to detect moonquakes. However, the lander's resonant vibration changes a lot under extreme temperature variations during the polar days and nights, which generates "mechanical noise" to contaminate real seismic signals. Besides, the solar panel mounted on the upper side of the lander body has to frequently rotate for the power supply, which may change the fundamental mode of lander's resonance. However, these influences remain unknown thus would leave great risk on the data mining. Here, we construct a numerical model of the lander and perform numerical simulations under a possible range of the temperature variation and solar panel position. We found that the lander's fundamental frequency drifts significantly between 0.64 Hz and 0.87 Hz under combined solar panel rotation and thermal cycling. This frequency range highly overlaps with that of real moonquake signals thus would contaminate the data. Our study provides a critical guide for scientists to recognize this specific frequency-varying "noise" and avoid potential misinterpretations on these data, which is crucial for an accurate detection of the Moon's interior structures.

**Abstract**

The Chang'e-7 (CE-7) mission will deploy the first seismometer at the lunar south pole to detect moonquakes and probe lunar interior structures in 2026 winter. However, the lander's vibration response to the extreme temperature cycles of the polar environment remains unclear, complicating the analysis of noise sources in seismic records. Here, we developed a high-fidelity finite-element model of the CE-7 lander to characterize its resonant behavior under the coupled influence of solar panel rotation and extreme thermal variations. Numerical results reveal that the lander's fundamental frequency (~0.76 Hz) at room temperature drifts significantly between 0.64 Hz and 0.87 Hz when the outside temperature varies from –180 to +80 °C. This frequency drift is primarily driven by thermally induced stiffness changes in the solar array supporting bracket, whereas geometric reconfiguration due to rotation plays a secondary role. Crucially, this resonance band directly overlaps with the primary seismic observation window (usually <1.0 Hz). Sensitivity analysis further confirms that the fundamental mode remains structurally robust despite material property uncertainties. These findings establish an essential theoretical baseline for identifying and filtering lander-induced resonant noise, which will be immediately applicable upon the acquisition of the first in-situ seismic datasets from the south pole of the Moon, ensuring the high fidelity of accurate lunar interior detection.

## 1 Introduction

Planetary seismology is the primary method for constraining the interior structure and thermal evolution of celestial bodies [Lognonné, 2005; Nunn et al., 2020]. Following the pioneering Apollo lunar seismic experiments and the recent InSight mission to Mars, China's Chang'e-7 (CE-7) mission is scheduled to deploy a broadband seismometer at the lunar south pole (Figure 1) to investigate the Moon's interior structures and seismicity [Zhang et al., 2022], and the distribution of volatiles such as water ice [Hurley et al., 2012; Wang et al., 2024; Zhang et al., 2022; Zhang et al., 2026; Zou et al., 2020]. Unlike terrestrial observatories where sensors are buried under the ground surface, planetary seismometers are frequently deployed directly on the ground, usually close to or even on the lander platform due to engineering constraints [Lognonné et al., 2019; Panning et al., 2020; Rosa et al., 2012]. Consequently, the lander structure functions as a complex mechanical filter and a potential source of parasitic noise, directly modulating the seismic signals recorded by the instrument [Compaire et al., 2021; Murdoch et al., 2018; Pakkathillam et al., 2025].

In planetary seismology, the coupling between extreme environments and lander dynamics constitutes a primary source of interference. Early experience from the Viking 2 mission [Lorenz, 2016] highlighted this vulnerability: its deck-mounted seismometer was overwhelmed by wind-induced lander resonances, which effectively masked tectonic signals [Anderson et al., 1977]. Even the more advanced InSight mission [Lognonné et al., 2019], which deployed its sensor meters away from the lander using a robotic arm and utilized wind and thermal shields, still recorded lander resonant signals [Ceylan et al., 2021; Dahmen et al., 2021; Zhang et al., 2023a] (Figure 1). Unless these structural vibrations are accurately identified and filtered, they risk being misinterpreted as genuine seismic signals. However, unlike the wind-dominated Martian environment [Murdoch et al., 2017a, 2017b], the lunar south pole presents a distinct set of environmental challenges driven not by atmospheric conditions, but by extreme thermal conditions [Paige et al., 2010; Williams et al., 2017], which ranges from –190 to +120 °C, much more severe than that at low latitudes on Mars (usually –100 to 20°C) [Dahmen et al., 2021; Zhang et al., 2023b].

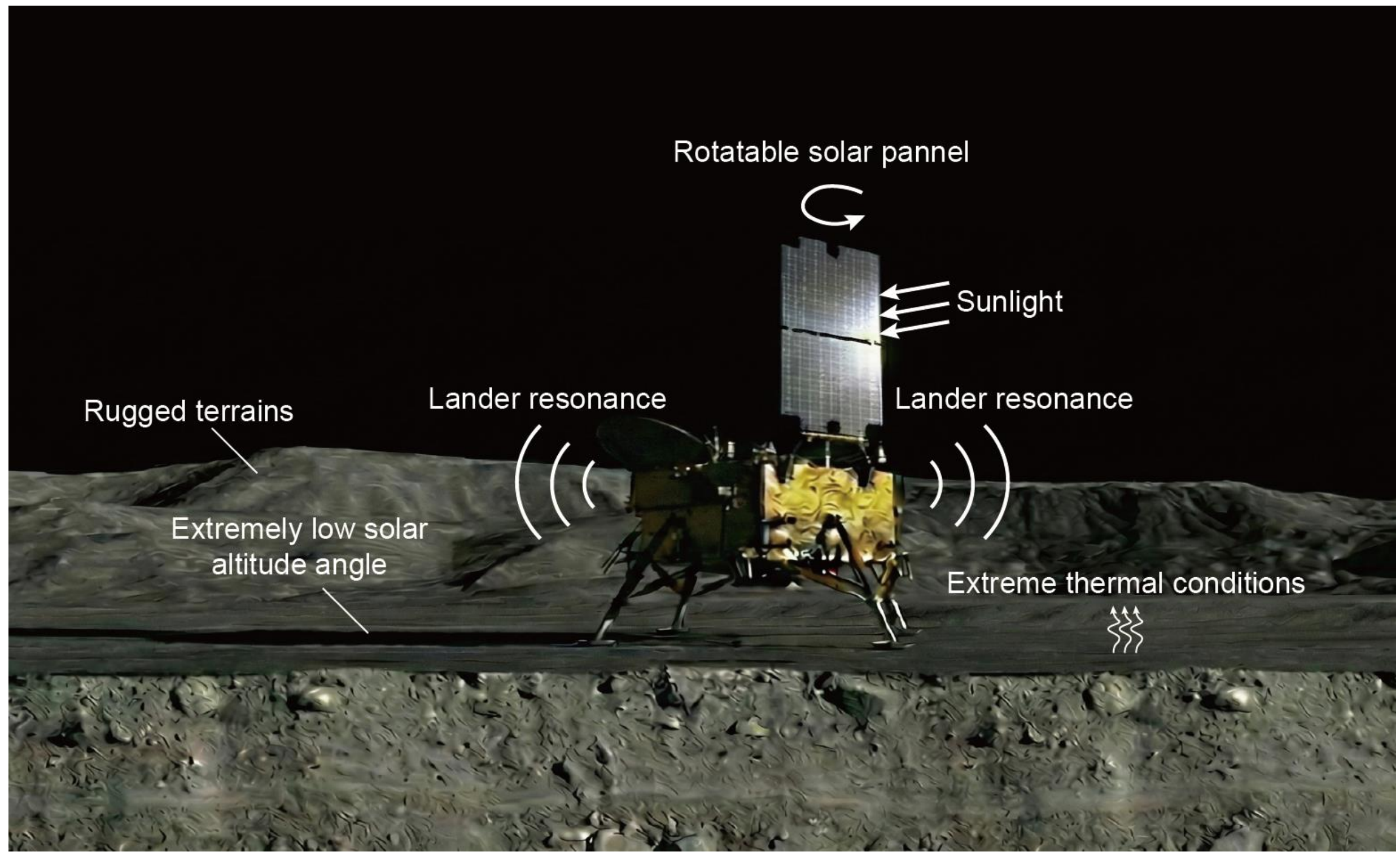

**Figure 1. Schematic representation of the CE-7 lander operating under the extreme environmental conditions of the lunar south pole**. The unique polar environment is characterized by an extremely low solar altitude angle, rugged terrains, and extreme thermal conditions with extensive shaded zones. To maintain the power supply, the large rotatable solar panel must be frequently tuned to track the sunlight. The coupling of these severe thermal gradients and the operational rotation of the solar panel excites structural vibrations, which would modulate the lander resonance. This mechanical noise constitutes a primary source of interference for in-situ seismic observations.

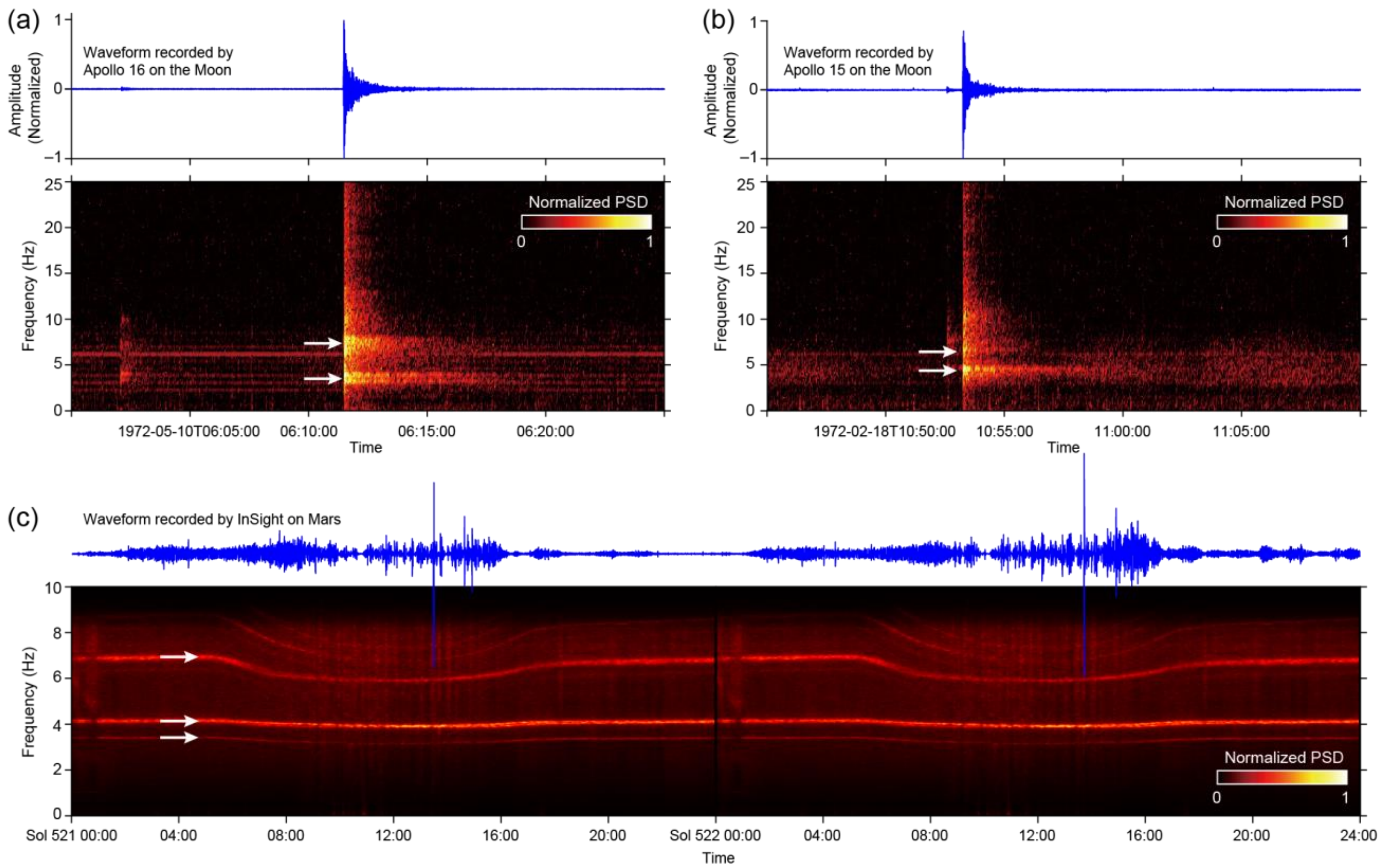

**Figure 2. Observational evidence of planetary lander structural resonances. (a)** Apollo 16 (Station 16 SPZ) seismic record of a moonquake event on May 10, 1972, with the corresponding spectrogram highlighting clear structural vibration signatures. **(b)** Apollo 15 (Station 15 SPZ) seismic record of a moonquake event on February 18, 1972, with the corresponding spectrogram highlighting clear structural vibration signatures. (**c**) Time-frequency spectrogram of a seismic event recorded by the InSight mission on Mars (Sols 521 and 522), showing distinct resonance bands originating from the lander's structural modes.

Archival data from Apollo missions confirmed this susceptibility, recording thermal moonquakes that contained distinct structural resonances of the Lunar Module driven by sunrise and sunset heating [Duennebier and Sutton, 1974; Liu et al., 2024; Onodera, 2024; Tamama et al., 2025] (Figure 2). Extensive re-evaluations of these early long-period and short-period event catalogs further emphasize the necessity of separating structural noise from genuine seismic events [Bulow et al., 2005; Nakamura et al., 1981]. However, the CE-7 mission faces an even greater challenge since it will land near the south pole: the lunar surface may experience an extreme thermal fluctuation up to 300°C during polar days and nights [Heiken et al., 1991]. Such a strong thermal stress is exacerbated when the solar elevation angle is extremely low (~3°) [Bussey et al., 1999, 2001; Litaker et al., 2025; Mazarico et al., 2011; Wang et al., 2024; Wei et al., 2023]. Consequently, the lander is exposed to sustained polar day-night cycles, resulting in long-term thermal deformation and significant variations in material properties due to the extreme low-temperature conditions. It is worth noting that all previous planetary seismometers have been deployed in low-latitude regions characterized by favorable solar illumination. In such environments, the diurnal solar trajectory follows a standard east-to-west arc, with direct overhead illumination at local noon. This geometry allows flat, horizontally mounted solar panels to efficiently sustain operations through typical day-night cycles spanning hours (e.g., on Mars) or 14 Earth days (e.g., on the Moon). However, at the lunar south pole, the duration of polar days and nights extend to over a hundred days. Coupled with an extremely low solar elevation angle, the daytime illumination is fundamentally altered into a slow, grazing azimuthal rotation, leaving the lander's top deck almost entirely unilluminated. Conversely, the prolonged polar night imposes extended periods without any solar power generation. This unprecedented and highly unique thermal environment poses immense challenges for the lander's structural and thermal design, inevitably inducing complex thermo-elastic deformations and material stiffness variations. Consequently, it remains unknown how these time-varying thermal loading modulates the lander's structural resonances and potentially interferes with seismic observations.

Reproducing the lunar polar environment for full-scale ground testing presents significant technical challenges. Due to the large dimensions (>5 m in diameter) of the deployed CE-7 lander [Wang et al., 2024; Zou et al., 2020], simultaneously simulating high vacuum and the extreme temperature range exceeds the capabilities of standard thermal-vacuum chambers. Consequently, numerical simulation serves as the primary method to predict the lander's dynamic response [Chen et al., 2014; Thornton & Kim, 1993]. The fundamental scientific problem is to quantify the frequency drift of the lander's modes under these coupled environmental extremes, establishing a theoretical baseline for distinguishing mechanical noise from seismic signals.

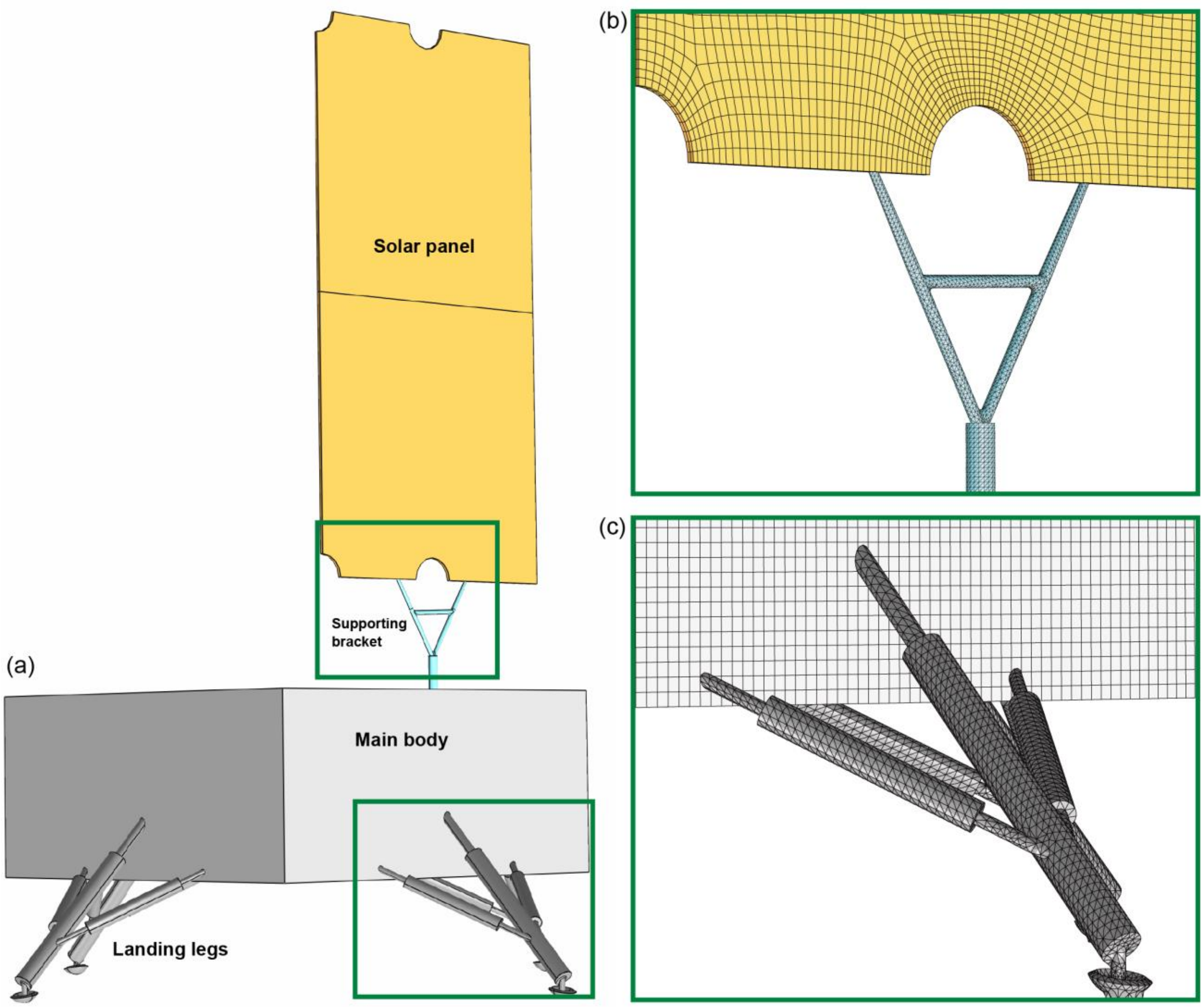

**Figure 3**. The finite element model and mesh details of the numerical CE-7 lander. (**a**) Global geometric model of the lander, illustrating the main body, the large-scale solar panel, and the landing legs. (**b**) Close-up view of the finite element mesh for the solar panel with its supporting bracket. (**c**) Detailed mesh configuration of the landing leg support structure. To ensure a high-fidelity simulation of the dynamic response, all components are discretized using solid elements.

To address these challenges, this study develops a high-fidelity finite-element model of the CE-7 lander to investigate its dynamic response under the specific environmental conditions of the lunar south pole. We systematically analyze the coupling effects between the continuous rotation of the solar arrays during the solar days and a typical annually thermal cycling (–180 °C to 80 °C). Specifically, the model incorporates temperature-dependent material properties to simulate the stiffness evolution of the structure during the polar day-night transitions [Zhang et al., 2023a]. Furthermore, a sensitivity analysis is conducted [Haftka & Adelman, 1989; Mottershead & Friswell, 1993] to determine the influence of key structural components on the migration of the lander's fundamental resonance frequencies.

## 2. Modeling of the CE-7 lander and modal analysis under sun-tracking conditions

We focus on low-frequency signals (below 20 Hz, specifically <1 Hz) that are critical for lunar interior detection. To balance computational efficiency with the physical accuracy required for macroscopic dynamic analysis, the finite-element model was constructed with specific structural idealizations tailored to our primary scientific objective, following Zhang et al. (2023a). First, the model focuses on the primary load-bearing and low-frequency structural components, as these directly couple with the primary seismic observation band (usually <1.0 Hz). High-frequency sub-structures were strategically simplified to isolate the global resonance behaviors. Second, the global structure was discretized using solid elements, and the complex composite parts were treated as homogenized, isotropic materials. This equivalent approach efficiently and effectively captures the overall mass distribution and global stiffness essential for fundamental mode analysis. Third, while the actual CE-7 lander accommodates diverse scientific payloads [Wang et al., 2024; Zou et al., 2020] that create asymmetric mass and thermal insulation configurations (e.g., across the +X and –Y quadrants), our baseline model adopts a generalized, symmetric approximation for the main body. This representative setup allows us to rigorously isolate and quantify the fundamental thermo-structural coupling mechanisms driven by the solar wing rotation and extreme polar temperatures, without being confounded by highly specific, localized payload asymmetries. As a result, we retain only the key macro structures (Figure 3): the solar panel (over 3 m high and nearly 2 m wide), the supporting bracket, the main body, and the landing legs. The entire assembly is modeled using solid elements, with material parameters evaluated based on realistic engineering data (listed in Table 1).

We performed a finite element analysis to investigate the lander's self-vibration characteristics following the method proposed for InSight Mission by [Zhang et al., 2023a]. Our results indicate that the first six modes are dominated by the flexible solar panel, with a fundamental frequency of 0.76 Hz (Figure 4). Crucially, this frequency falls within the sensitive bandwidth for lunar interior structure analysis, suggesting that structural resonances could interfere with seismic data interpretation. Regarding the operational phase, since the solar panel must frequently rotate to track the sun for power supply, we modeled this dynamic process explicitly. Specifically, to simulate the sun-tracking effect under different solar illumination conditions, we incrementally rotated the solar panel from 0˚ to 180˚ relative to the fixed lander body, while keeping all other parts unchanged, following the basic idea on the design of CE-7 lander [Wang et al., 2024; Zou et al., 2020]. The simulated results (Figure 4) reveal that while this rotation alters the vibration direction, the natural frequencies remain remarkably stable across the entire spectrum. Specifically, the fundamental frequency (Mode 1) is locked at 0.76 Hz. This stability extends to higher-order bands: the torsional Mode 3 (~4.00 Hz) and the complex coupled Mode 5 (~14.68 Hz) exhibit negligible fluctuations of less than 2%, with no mode crossing observed. This confirms that the lander's dynamic signature is determined by the solar panel's intrinsic stiffness, independent of its tracking orientation. Most previous planetary landers tend to place the solar panels horizontally, which would cause an evident contrast of total illumination intensity on the solar panel between the morning/evening and noon times, if the solar panels are not intentionally reoriented. In contrast, the solar panel is placed vertically on the top of CE-7 lander, which has to operate frequently or continuously during the polar days to capture sufficient sunlight; thus, its total illumination intensity is generally stable in the daytime. Therefore, we conclude that the resonance of the CE-7 lander is slightly or even not influenced by its solar panel, contrary to the InSight lander.

## 3. Impact of thermal environment on vibration characteristics

Considering the extreme thermal environment of the landing site at the lunar south pole, characterized by significant temperature fluctuations, we performed a finite element analysis to quantify the impact on the lander's vibration characteristics. Following the methodology of Zhang et al. (2023a), we simulated these thermal effects by adjusting material stiffness—specifically, modeling the 'softening' behavior at high temperatures and the 'hardening' effect at low temperatures. This approach is physically grounded in experimental studies, as confirmed experimental studies showing that the elastic modulus of metallic materials increases as temperature decreases [Çolakoğlu, 2007; Yu et al., 2014; Han et al., 2020; Zhanget al., 2023a]. The simulation results reveal that while the mode shapes remain structurally stable, the natural frequencies exhibit distinct shifts due to these thermal-elastic effects. Under extreme cold (−180℃, Figure 5b), material stiffening raises the fundamental frequency (Mode 1) from 0.76 Hz to 0.80

Hz and significantly increases the higher-order Mode 6 from 18.02 Hz to 18.92 Hz. Conversely, at high temperatures (80℃, Figure 5c), material softening leads to a slight frequency drop (e.g., Mode 1 to 0.75 Hz). This confirms that the solar panel's structural integrity is robust and predictable across the full lunar temperature range.

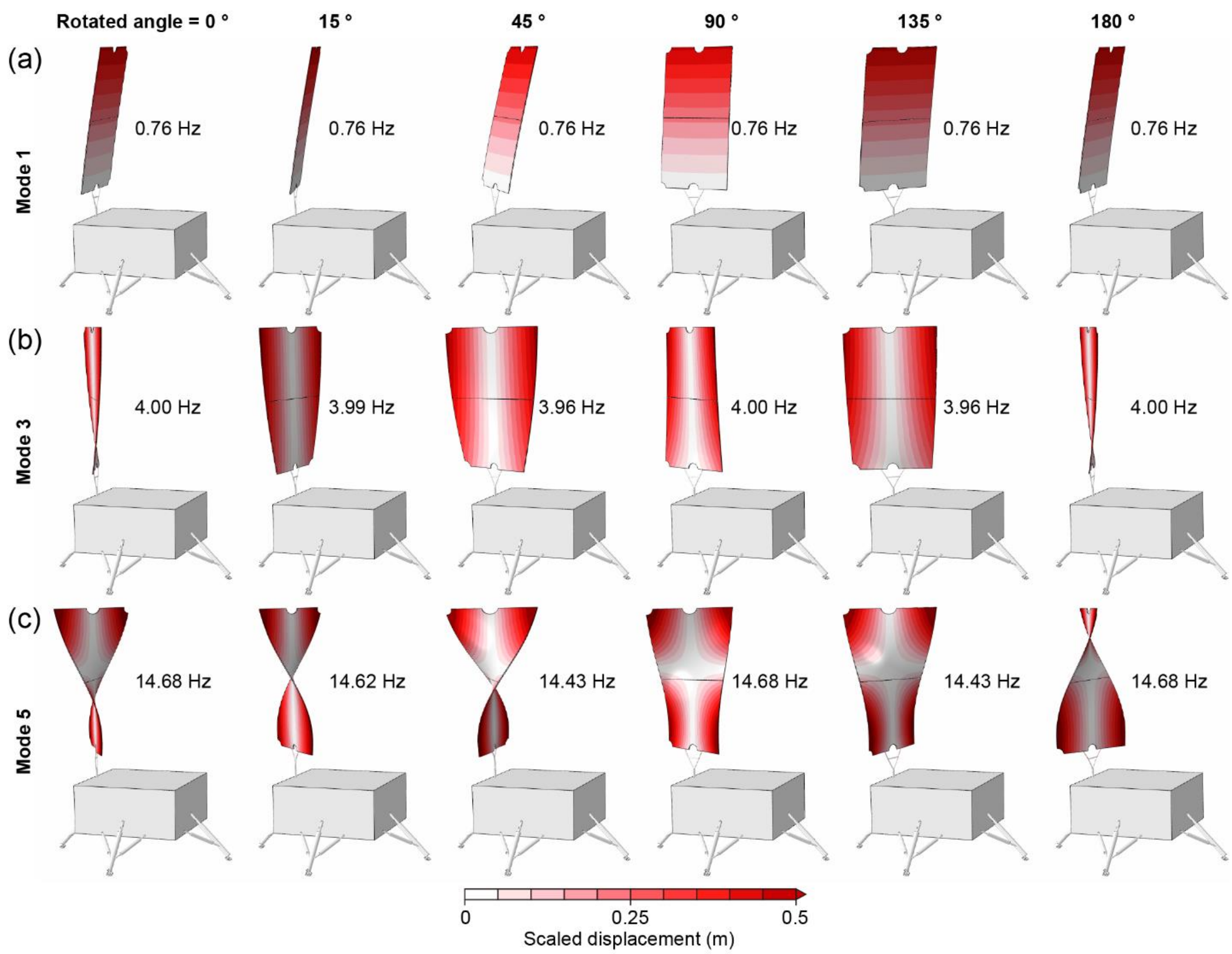

**Figure 4. Evolution of the key resonant modes of the CE-7 lander during simulated sun-tracking operations.** The solar panel is rotated incrementally from 0˚ to 180˚ relative to the lander body. (**a**) Mode 1 (Fundamental Bending); (**b**) Mode 3 (Torsional/Coupled); and (**c**) Mode 5 (Higher-order Bending/Twisting). The color contours represent the modal displacement field. The values of resonance frequency are presented near the solar panel.

**Table 1. Material parameters of the simplified model**

| Parts | Density (kg/m$^3$) | E (Pa) | | | Poisson's ratio |
|---|---|---|---|---|---|
| | | T=20℃ | T=-180℃ | T=80℃ | |
| SolarWing | 200 | 1.65E+10 | 1.82E+10 | 1.60E+10 | 0.3 |
| WingsSupport | 4430 | 1.14E+11 | 1.25E+11 | 1.11E+11 | 0.34 |
| MainBody | 135 | 3.15E+09 | 3.50E+09 | 3.05E+09 | 0.3 |
| Feet | 600 | 2.63E+10 | 2.92E+10 | 2.54E+10 | 0.33 |

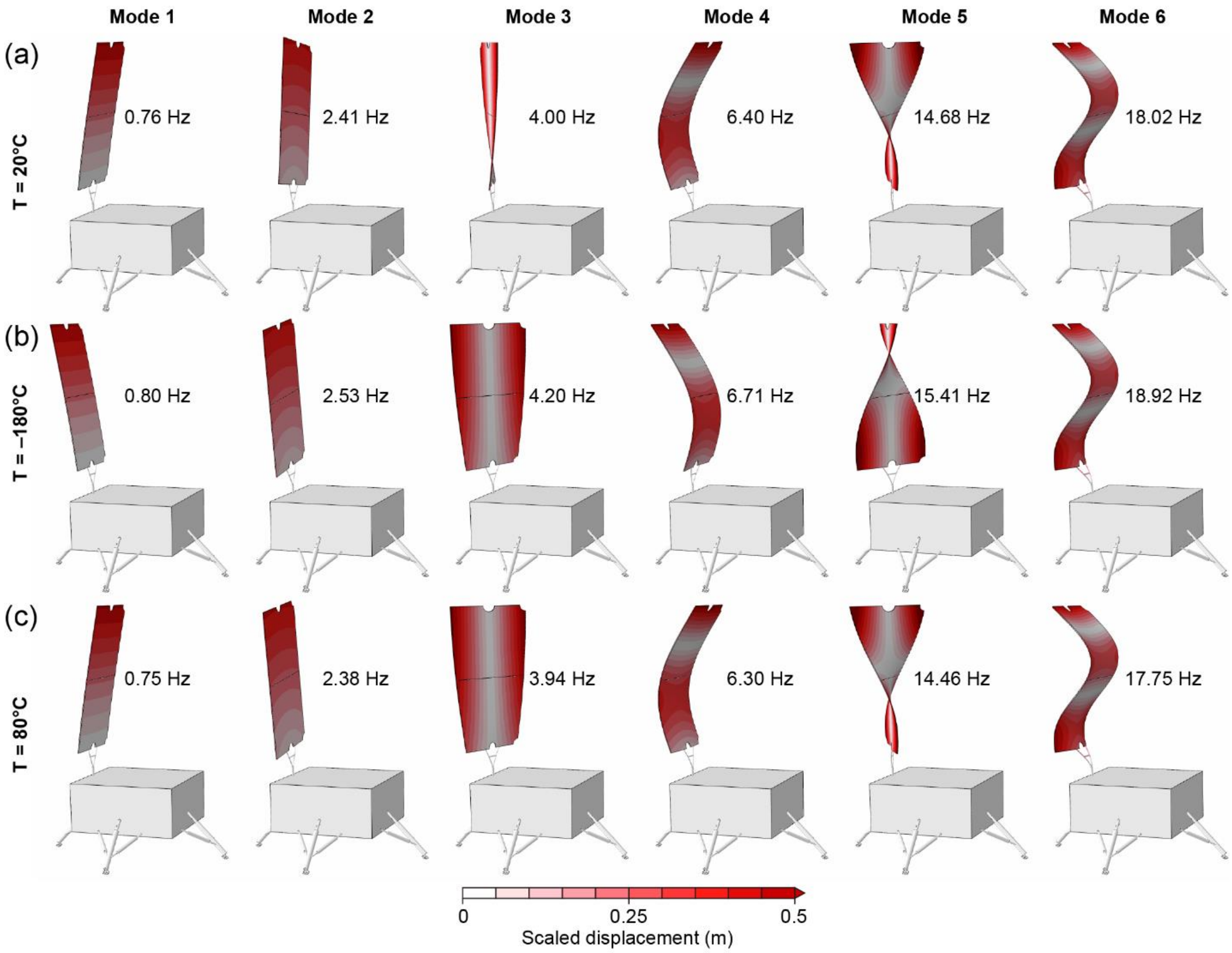

**Figure 5. Thermal sensitivity analysis of the CE-7 lander's vibration modes.** The evolution of the first six mode shapes and natural frequencies is presented under three distinct thermal conditions: (**a**) Reference temperature (T = 20°C); (**b**) Extreme low temperature (T = –180°C); and (**c**) High temperature (T = 80°C). The specific natural frequency value is annotated next to each mode shape to quantify the thermal effect.

Building on the previous thermal analysis (Figure 5), which identified the solar panel as the primary flexible component, we further investigated the impact of material property variations. As reported by Zhang et al. (2023), continuous solar heating can significantly alter the stiffness of solar panels, leading to elastic modulus fluctuations of up to 30%. Consequently, to simulate how extreme temperature variations (up to 260°C) modulate structural stiffness through thermally induced material softening and stiffening, we varied the elastic modulus ($E$) within a range of ±30% for the solar panel, the supporting bracket, and the solar panel combined with the supporting bracket, respectively (Figure 6). As shown in Figure 6, variations in the supporting bracket stiffness (Figure 6d) induce a frequency drift ranging from 0.69 Hz to 0.82 Hz. In comparison, the solar panel (Figure 6b) has a smaller influence, with the frequency shifting between 0.74 Hz and 0.81 Hz, indicating that the supporting bracket acts as the primary stiffness bottleneck for the fundamental mode. When both parts (solar panel combined with the supporting bracket) are varied simultaneously to simulate extreme environmental conditions (Figure 6f), the fundamental frequency exhibits the most significant drift, spanning from 0.64 Hz to 0.87 Hz under the ±30% variation bounds. Furthermore, higher-order modes (Modes 4–6) demonstrate considerably higher sensitivity to stiffness changes than the fundamental mode, suggesting that material property variations become increasingly critical for higher-frequency structural dynamics.

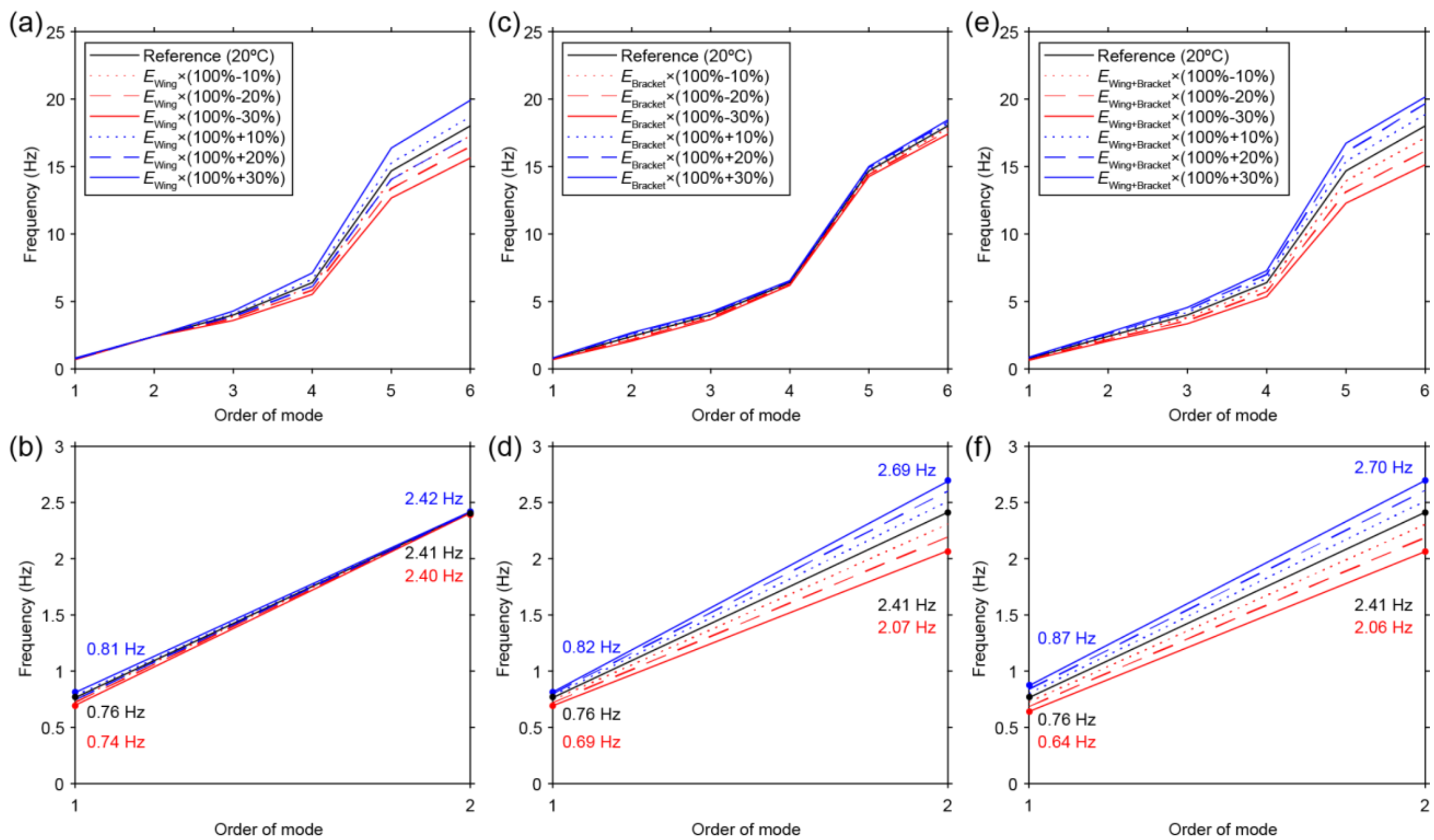

**Figure 6. Sensitivity analysis of the lander's natural frequencies to material stiffness variations. (a)** The plot illustrates the frequency shifts for the first six modes when the elastic modulus ($E$) of the solar panel is perturbed by ±10%, ±20%, and ±30% relative to the reference state (black line). The blue and red dashed lines represent stiffness increases and decreases, respectively. (**b**) The zoomed-in view of the first two modes with frequency values marked. (**c-d**) The same as a-b but for the supporting bracket. (**e-f**) The same as a-b but for the solar panel combined with supporting bracket.

## 4. Discussion

Although the CE-7 seismometer operates during polar days with sun-tracking capabilities, the unique topography of the lunar south pole introduces complex thermal constraints [Paige et al., 2010; Williams et al., 2017]. Due to the extremely low solar elevation angle (<3∘), the lander is subject to intermittent shadowing caused by local terrain features even during operational windows [Bussey et al., 2010; Mazarico et al., 2011; Wei et al., 2023]. This results in rapid, severe thermal fluctuations, forcing the solar array to transition between high-temperature states (+80°C) and cryogenic conditions (−180°C). The trend of resonance frequency shifts induced by such rapid temperature variations is likely consistent with the processes demonstrated in this study. However, we cannot rule out the possibility that these abrupt heating and cooling cycles may generate transient seismic 'glitches' triggered by the sudden thermal expansion and contraction of the lander structure. While our current methodology cannot directly predict the occurrence of these transient phenomena, lessons from the Mars InSight mission indicate that our modal analysis could be integrated to help identify and mitigate such thermal glitches in the actual seismic records (Xu et al., 2022).

To fully contextualize the dynamic behavior of the CE-7 lander, it is necessary to contrast it with the Mars InSight mission, highlighting how latitude, illumination geometry, and structural design fundamentally alter their thermo-structural responses. First, regarding environmental excitation, the Martian atmosphere provides a continuous wind-induced driving force that constantly agitates InSight's solar panels—a mechanism entirely absents in the lunar high-vacuum environment [Lorenz, 2016; Murdoch et al., 2017]. Second, the two landers experience entirely different solar heating patterns. InSight operated at mid-to-low Martian latitudes, featuring two solar arrays symmetrically distributed along the east-west axis and facing upwards. Consequently, the InSight lander received relatively uniform, top-down solar irradiation during the day, leading to symmetric but drastic thermal expansion and contraction between day and night [Pakkathillam et al., 2025; Zhang et al., 2023a]. In stark contrast, the CE-7 mission is targeted at the lunar south pole, where the extremely low solar elevation angle (~3°) results in highly inclined, grazing illumination [Gläser et al., 2014; Noda et al., 2008]. Unlike InSight's uniform top-down heating, this low-angle sunlight creates a highly uneven thermal distribution across the CE-7 lander. The grazing sunlight intensively heats only one or two specific sides of the main body, leaving the opposite sides in deep cryogenic shadow. Furthermore, unlike InSight's symmetric design, the CE-7 lander body is highly asymmetric, with varying mass distributions and thermal insulation configurations across its different sides. As the inclined sunlight slowly sweeps around the horizon during the prolonged polar day, it sequentially heats these physically distinct sides. This highly directional and uneven thermal input is the fundamental mechanism driving the complex, orientation-dependent deformations and significant frequency drifts observed in our model. Meanwhile, the CE-7's single, distantly mounted solar panel cleverly bypasses this severe thermal cycling by frequently or continuously rotating to track the sun, thereby maintaining a stable illumination [Thornton et al., 1995; Zhu et al., 2025].

A fundamental limitation of this pre-flight analysis is the absence of in-situ observational constraints from the lunar south pole. Consequently, the current model relies on theoretical material parameters and ground-based estimates, which may deviate from the actual operational state. Once the CE-7 seismometer collects in-situ data, the lander's real self-vibration characteristics can be identified. These measurements will serve as a "ground truth" to optimize key model parameters—particularly the material properties of the flexible solar panels—thereby precisely calibrating the specific vibration modes [Mottershead & Friswell, 1993, Kreider & Arya, 2024]. This updated model will provide a critical basis for filtering out mechanical resonance noise, ensuring the high fidelity of seismic data required for the accurate inversion of the Moon's interior structure.

## 5. Conclusions

This study characterizes the dynamic response of the CE-7 lander under the extreme environmental conditions of the lunar south pole, providing a baseline for distinguishing mechanical artifacts from moonquakes. Our finite element analysis reveals that the lander's fundamental frequency is not static but drifts significantly between 0.64 Hz and 0.87 Hz. This wide variation is primarily driven by the drastic annual temperature fluctuations (–180°C to +80°C), which modulate the material stiffness of the structure. Furthermore, unlike the wind-driven and symmetric thermal cycling observed on Martian landers, CE-7's thermo-structural response is uniquely governed by highly directional, grazing polar sunlight sequentially heating its asymmetric main body, while its sun-tracking solar panel maintains relative thermal stability. Crucially, sensitivity analysis identifies the supporting bracket of the solar array as the primary stiffness bottleneck governing the resonance mode. Its frequency band directly overlaps with the primary seismic observation window (usually <1.0 Hz), highlighting the necessity of time-varying noise filtering. Most importantly, our results establish a quantitative trend of frequency drift as a function of solar illumination azimuth. This orientation-dependent signature provides a critical diagnostic criterion for recognizing temperature-induced lander artifacts in actual seismic records, strictly preventing the misinterpretation of mechanical noise as genuine lunar seismic events. Ultimately, this pre-flight numerical analysis establishes a predictive framework for isolating and filtering lander-induced resonances. This capability will be immediately deployable once the CE-7 mission returns its first seismic datasets from the lunar south pole. By enabling the precise subtraction of these time-varying mechanical artifacts, our results safeguard the high data fidelity essential for accurate imaging and inversion of the Moon's deep interior.

## Acknowledgments

This work is supported by the National Natural Science Foundation of China (NSFC) Grants No. 42325406. The authors thank Yanan Zhang for helpful discussions regarding the finite element modeling.

## Data Availability Statement

The Apollo seismic data were retrieved from the International Federation of Digital Seismograph Networks (Nunn, 2022). The InSight seismic data were downloaded from NASA's Planetary Data System (PDS, https://pds-geosciences.wustl.edu/missions/insight/seis.htm). The ObsPy Python toolbox (Beyreuther et al., 2010) for seismic data processing is used in this study. The modal analysis was performed using the FEM software Abaqus.

## Conflict of Interest Disclosure

The authors declare there are no conflicts of interest in this manuscript.